\documentstyle[preprint,aps,epsfig]{revtex}
\tightenlines
\begin{document}

\preprint{
\vbox{
\hbox{ADP-00-21/T404}
\hbox{JLAB-THY-00-11}
}}

\title{Dynamical Symmetry Breaking in the Sea of the Nucleon}

\author{A.W.~Thomas$^1$, W.~Melnitchouk$^{1,2}$ and F.M.~Steffens$^3$}
\address{$^1$	Special Research Centre for the
		Subatomic Structure of Matter,
		and Department of Physics and Mathematical Physics,
		Adelaide University, Adelaide SA 5005,
		Australia}
\address{$^2$	Jefferson Lab,
		12000 Jefferson Avenue,
		Newport News, VA 23606}
\address{$^3$	Instituto de Fisica -- USP,
		C.P. 66 318, 05315-970,
		Sao Paulo, Brazil}

\maketitle

\begin{abstract}
We derive the non-analytic chiral behavior of the flavor asymmetry
$\bar{d} - \bar{u}$.
Such behavior is a unique characteristic of Goldstone boson loops in
chiral theories, including QCD, and establishes the unambiguous role
played by the Goldstone boson cloud in the sea of the proton.
Generalizing the results to the SU(3) sector, we show that strange
chiral loops require that the $s - \bar s$ distribution be non-zero.
\end{abstract}

\vspace*{1cm}

Understanding the role of dynamical chiral symmetry breaking in hadron
structure is one of the central problems in strong interaction physics.
On very general grounds one can show that Goldstone boson loops make
significant contributions to hadronic properties such as charge
distributions and magnetic moments.
For example, in the chiral limit the charge radii of the proton and
neutron are known to diverge as $\ln m_\pi$ \cite{RADII}.  
Such non-analytic behavior as a function of quark mass (recall that
$m_\pi^2 \sim \overline{m} = (m_u + m_d)/2$) is a unique characteristic
of Goldstone boson loops.

Historically the focus for the role of dynamical chiral symmetry breaking 
in hadron structure has been on low energy properties such as masses and
electromagnetic form factors.
On the other hand, the possibility of an excess of $\bar{d}$ over
$\bar{u}$ quarks in the proton was predicted on the basis of the
nucleon's pion cloud \cite{PION83}.
Since the experimental verification that indeed $\bar{d} > \bar{u}$
\cite{EXPT}, many analyses have been presented in which the pion cloud
is a major source of the asymmetry \cite{EARLY,CLOUD}.
Yet there has so far been no rigorous connection established between
these models and the chiral properties of QCD.
As a result, the fundamental importance of the pion contribution to
the flavor asymmetry has not been universally appreciated.

In this Letter, we establish for the first time the non-analytic chiral
behavior of $\bar{d} - \bar{u}$ and hence the unambiguous role of the
Goldstone boson cloud in the flavor asymmetry of the nucleon sea.
It turns out that the leading non-analytic (LNA) behavior of the excess
number of $\bar{d}$ over $\bar{u}$ quarks in the proton has a chiral
behavior typical of loop expansions in chiral effective theories,
such as chiral perturbation theory \cite{CHIPT}.
Specifically, we find:
\begin{eqnarray}
\label{result}
\left( \overline{D} - \overline{U} \right)_{\rm LNA}^{(0)}
&\equiv& \int_0^1 dx \left( \bar{d}(x) - \bar{u}(x) \right)_{\rm LNA}
\ =\ { 2 g_A^2 \over (4 \pi f_\pi)^2 }\
m_\pi^2  \log (m_\pi^2/\mu^2) ,
\end{eqnarray}
where $g_A$ is the axial charge of the nucleon 
(understood to be taken in the chiral SU(2) limit,
$\overline{m} \to 0$), and $\mu$ is a mass parameter.
This result also generalizes to higher moments, each of which has a
non-analytic component, so that the $\bar d - \bar u$ distribution
itself, as a function of $x$, has a model-independent, LNA component.
The presence of non-analytic terms indicates that Goldstone bosons
play a role which cannot be cancelled by any other physical process
(except by chance at a particular value of $m_\pi$).
Such insight is vital when it comes to building models and developing
physical understanding of a system.

In deep-inelastic scattering the one-pion loop contribution to the
$n$-th moment of the $\bar d(x) - \bar u(x)$ difference is given by
\cite{PION83,CLOUD}:
\begin{eqnarray}
\label{d_umom}
\left( \overline D - \overline U \right)^{(n)}
&=& \int_0^1 dx\ x^n\ (\bar d(x) - \bar u(x) )\
 =\ {2 \over 3}\ V_\pi^{(n)}\ \cdot f_{\pi N}^{(n)}\ ,
\end{eqnarray}
where $V_\pi^{(n)}$ is the $n$-th moment of the valence pion structure
function\footnote{The assumption implicit in the appearance of the pion
valence distribution is that the sea of the pion is flavor symmetric.
The generalization to the case where this is not so is straightforward,
but this contribution would be confined to very small values of
Bjorken $x$.},
and $f_{\pi N}^{(n)}$ is the $n$-th moment of the pion distribution
function in the nucleon (or the $N \to \pi N$ splitting function):
\begin{eqnarray}
f_{\pi N}^{(n)} &=& \int_0^1 dy\ y^n\ f_{\pi N}(y)\ .
\end{eqnarray}
The momentum dependence of the pion distribution function is
given by \cite{CLOUD,SULL}:
\begin{eqnarray}
\label{fypiN}
f_{\pi N}(y) &=& \left( { 3 g_{\pi NN}^2 \over 16 \pi^2 } \right)
y \int_{t_{min}}^{\mu^2}\ dt\ { t \over (t + m_\pi^2)^2 }\ ,
\end{eqnarray}
where $t = -k_\mu k^\mu$ ($k_\mu$ is the four-momentum of the pion),
with a minimum value $t_{min} = M^2 y^2/(1-y)$ determined from the
on-shell condition for the recoil nucleon, and $g_{\pi NN}$ is the
$\pi NN$ coupling constant.
Since the non-analytic structure of pion loops does not depend on
the short-distance behavior, we have for simplicity introduced an
ultra-violet cut-off, $\mu$, to regulate the integral in
Eq.~(\ref{fypiN}).
One could have equally well used a form factor for the $\pi NN$ vertex,
or a more elaborate regularization procedure.

It is vital to understand that this contribution to $\bar d - \bar u$ is
a leading twist contribution to the structure function of the nucleon.
The hard scattering involves the constituents of the pion itself, while
the momentum of the pion is typical of those met in chiral models of
nucleon structure, namely a few hundred MeV/c.
The fact that the momentum associated with the pion is low is the reason
one can discuss the LNA structure of $\bar d - \bar u$.
There may, of course, be other terms which contribute to the physical
difference between $\bar d$ and $\bar u$, which cannot be expressed in
the factorized form of Eq.~(\ref{d_umom}), such as interactions of the
spectator quark in the pion with the recoil nucleon.
However, the LNA behavior of $\bar d - \bar u$ is entirely determined
by the one-pion loop and cannot be altered by such contributions.

Taking the $n$-th moment of the distribution in Eq.~(\ref{fypiN}),
the LNA chiral log contribution from a pion loop is:
\begin{equation}
\label{fpiN_LNA}
\left. f_{\pi N}^{(n)} \right|_{\rm LNA} \ =\
\left( 3 M^2 g_A^2 / (4 \pi f_\pi)^2 \right) \times
\left\{
\begin{array}{l}
	(-1)^{n/2} ((n+4)/(2n+4))
	\left( m_\pi/M \right)^{n+2}
	\log (m_\pi^2/\mu^2)				\nonumber\\
	\hspace*{5cm} (n=0,2,4,\cdots) ,		\\
        (-1)^{(n+1)/2} ((n+5)/2)
	\left( m_\pi/M \right)^{n+3}
	\log (m_\pi^2/\mu^2)				\nonumber\\
	\hspace*{5cm} (n=1,3,5,\cdots) ,
\end{array}
\right.
\end{equation}
where the PCAC relation has been used to express the $\pi NN$ coupling
constant in terms of the axial charge $g_A$ (both $g_A$ and the nucleon
mass, $M$, are taken in the chiral SU(2) limit).
For the $n=0$ moment, conservation of baryon number requires that
$V_\pi^{(0)} = 1$, which leads directly to Eq.~(\ref{result}).
The LNA contributions to the $n > 0$ moments are suppressed in
the chiral limit by additional powers of $m_\pi^n$.
The scale dependence of $V_\pi^{(n)}$ for $n > 0$ introduces a $Q^2$
dependence into the higher moments of $\bar d - \bar u$.
In particular, the observed decrease with $Q^2$ of the $n > 0$ moments
of $\bar d - \bar u$ arises from the QCD evolution of the momentum
fraction carried by valence quarks in the pion ($V_\pi^{(n>0)}$).

Another contribution known to be important for nucleon structure is
that from the $\pi \Delta$ component of the nucleon wave function
\cite{CBM}.
For a proton initial state, the dominant Goldstone boson fluctuation is
$p \to \pi^- \Delta^{++}$, which leads to an excess of $\bar u$ over
$\bar d$.
The one-pion loop contribution to the $n$-th moment of $\bar d-\bar u$
from this process can be written in a similar form as Eq.~(\ref{d_umom}):
\begin{eqnarray}
\left( \overline D - \overline U \right)^{(n)}
&=& - {1 \over 3}\ V_\pi^{(n)}\ \cdot f_{\pi \Delta}^{(n)}\ ,
\end{eqnarray}
where $f_{\pi \Delta}^{(n)}$ is the $n$-th moment of the $\pi \Delta$
momentum distribution \cite{MST},
\begin{eqnarray}
f_{\pi \Delta}(y)
&=& \left( { 2 g_{\pi N \Delta}^2 \over 16 \pi^2 } \right)
y \int_{t_{min}}^{\mu^2}\ dt\
{ \left( t + (M_\Delta-M)^2 \right) \left( t + (M_\Delta+M)^2 \right)^2
\over 6 M_\Delta^2\ (t + m_\pi^2)^2 }\ ,
\end{eqnarray}
with $t_{min} = M^2 y^2/(1-y) + \Delta M^2\ y/(1-y)$, and
$\Delta M^2 = M_\Delta^2 - M^2$ (again the masses and the coupling
$g_{\pi N\Delta}$ are implicitly those in the chiral limit).
Evaluating the $n$-th moment of the $\pi \Delta$ distribution explicitly,
one finds the following LNA behavior:
\begin{eqnarray}
\left. f_{\pi \Delta}^{(n)} \right|_{\rm LNA}
&=&
{ 6 \over 25 } { g_A^2 \over (4 \pi f_\pi)^2 }
{ (M_\Delta+M)^2 \over M_\Delta^2 }
(-1)^n { m_\pi^{2n+2} \over \Delta M^{2n} }
\log (m_{\pi}^2/\mu^2) ,
\end{eqnarray}
where SU(6) symmetry has been used to relate $g_{\pi N \Delta}$ to $g_A$.

We stress that the current analysis aims only at establishing the
model-independent, chiral behavior of flavor asymmetries, without
necessarily trying to explain the entire asymmetries quantitatively.
It is interesting, nevertheless, to observe that with a mass scale
$\mu \sim 4\pi f_\pi \sim 1$~GeV, the magnitude of the LNA contribution
(at the physical pion mass) to the $n=0$ moment of $\bar d - \bar u$
is quite large --- of order 0.2, most of which comes from the $\pi N$
component.
For comparison, we recall that the latest experimental values for the
asymmetry $(\overline{D} - \overline{U})^{(0)}$ lie between
$\approx 0.1-0.15$ \cite{EXPT}.

In addition to $\pi\Delta$ intermediate states, contributions from other,
heavier baryons and mesons to the $\bar d-\bar u$ asymmetry have been
considered in meson cloud models \cite{HEAVY}.
Unlike the situation that we have explored for the (pseudo-Goldstone)
pion, however, there is no direct, model independent connection with
the chiral properties of QCD for mesons such as the $\rho$ and $\omega$.

One can generalize the preceding analysis to the flavor SU(3) sector by
considering the chiral behavior of the $s$ and $\overline s$ components
of the sea of the nucleon associated with kaon loops.
One finds that the non-trivial moments of the difference between the
$s$ and $\bar s$ distributions are non-analytic functions of
$\overline{m} + m_s$, with $m_s$ the strange quark mass.

As originally proposed by Signal and Thomas \cite{ST}, virtual kaon loops
are one possible source of non-perturbative strangeness in the nucleon
\cite{STRANGE}.
Unlike the case of SU(2) flavor asymmetry, however, where only the direct
coupling to the pion plays a role, both the kaon and hyperon (for example,
the $\Lambda$) carry non-zero strangeness and hence contribute to strange
observables.
Furthermore, the different momentum distributions of $\bar s$ quarks in
the kaon and $s$ quarks in the hyperon lead to different $s$ and $\bar s$
distributions as a function of $x$, as well as to non-zero values for
strange electromagnetic form factors \cite{STRANGE}.

The $n$-th moment of the $s-\bar s$ difference arising from a one-kaon
loop can be written \cite{ST}:
\begin{eqnarray}
\left( S - \bar S \right)^{(n)}\
 =\ \int_0^1 dx\ x^n\ \left( s(x) - \bar s(x) \right)
&=& V^{(n)}_{\Lambda} \cdot f_{\Lambda K}^{(n)}\
 -\ V^{(n)}_K \cdot f_{K \Lambda}^{(n)}\ ,
\end{eqnarray}
where $f_{K \Lambda}^{(n)}$ is the $n$-th moment of the $N \to K \Lambda$
splitting function:
\begin{eqnarray}
\label{fKL}
f_{K\Lambda}(y) &=& \left( { g_{KN\Lambda}^2 \over 16 \pi^2 } \right)
y\ \int_{t_{min}}^{\mu^2} dt\
	{ t + (M_\Lambda-M)^2 \over (t + m_K^2)^2 }\ ,
\end{eqnarray}
with $t_{min} = M^2\ y^2/(1-y) + \Delta M^2\ y/(1-y)$
and $\Delta M^2 = M_\Lambda^2-M^2$.
The corresponding moment of the $\Lambda$ distribution,
$f_{\Lambda K}^{(n)}$, can be evaluated from $f_{K \Lambda}^{(n)}$
through the symmetry relation between the splitting functions:
\begin{eqnarray}
\label{symm}
f_{\Lambda K}(y) &=& f_{K \Lambda}(1-y)\ .
\end{eqnarray}
Zero net strangeness in the nucleon implies the vanishing of the $n=0$
moment, $(S - \bar S)^{(0)} = 0$, which follows from Eq.~(\ref{symm}) and
strangeness number conservation, $V^{(0)}_{\Lambda} = V^{(0)}_K = 1$.
For higher moments, however, this is no longer the case, so that in
general $(S - \bar S)^{(n)}$ will be non-zero for $n > 0$.
In particular, the LNA components of the strange distributions will be
given by:
\begin{eqnarray}
\left. f_{K\Lambda}^{(n)} \right|_{\rm LNA}
&=& { 27 \over 25 } { M^2 g_A^2 \over (4 \pi f_\pi)^2 }
(M_\Lambda-M)^2 (-1)^n { m_K^{2n+2} \over \Delta M^{2n+4} }
\log (m_K^2/\mu^2)\ ,
\end{eqnarray}
where we have used SU(6) symmetry to relate $g_{KN\Lambda}$ to
$g_A/f_\pi$.
It is especially interesting to note that while the LNA part of the
$n$-th moment of $\bar s$ is of
order $m_K^{2n+2} \log m_K^2$, from Eq.(\ref{symm}) the LNA contribution
to the
$n$-th moment of $s$ is of order $m_K^2 \log m_K^2$.
As a consequence the entire $x$-dependence of $s(x) - \bar s(x)$ has
a LNA component of order $m_K^2 \log m_K^2$.
Since the LNA terms in the chiral expansion are model-independent,
and in general not cancelled by other contributions, this result
establishes the fact that the process of dynamical symmetry breaking in
QCD implies that the $s$ and $\bar s$ distributions must have a different
dependence on Bjorken $x$.

Experimental evidence for a strange--antistrange asymmetry is being
sought in deep-inelastic neutrino and antineutrino scattering experiments
by the CCFR Collaboration \cite{CCFR}.
At the present level of precision it is not possible to resolve the
asymmetry which, as we have shown, is expected on quite general grounds.
Nevertheless, it should be amenable to future measurements.

A similar analysis can also be performed for spin-dependent quark
distributions.
Although there will be no contribution to polarized asymmetries from
direct coupling to the Goldstone bosons, there will be indirect effects
associated with chiral loops via the interaction with the baryon which
accompanies the meson ``in the air''.
Such processes will renormalize the axial charge, for example, as well
as give rise to polarization of strange quarks.
Interestingly, Goldstone boson loops will not give rise to any flavor
asymmetries for spin-dependent antiquark distributions,
$\Delta \bar d - \Delta \bar u$, for which the only known source is Pauli
blocking effects in the proton \cite{PAULI}.

In summary, we have derived the leading non-analytic chiral behavior of
flavor asymmetries in the proton which are associated with Goldstone
boson loops.
These results establish the fact that the measurement of flavor
asymmetries in the nucleon sea reveals direct information on dynamical
chiral symmetry breaking in QCD.

\acknowledgements

We would like to thank W. Detmold and J. Goity for a careful reading of
the manuscript.
This work was supported by the Australian Research Council, U.S. DOE
contract \mbox{DE-AC05-84ER40150}, and FAPESP (96/7756-6, 98/2249-4).



\references

\bibitem{RADII}
M.A.B.~B\'eg and A.~Zepeda,
Phys. Rev. D 6, 2912 (1972);
%
J.~Gasser, M.E.~Sainio and A.~Svarc,
Nucl. Phys. B 307, 779 (1988);
%
D.B.~Leinweber and T.D.~Cohen,
Phys. Rev. D 47, 2147 (1993).

\bibitem{PION83}
A.W.~Thomas,
Phys. Lett. B 126, 97 (1983).

\bibitem{EXPT}
P.~Amaudraz et al.,
Phys. Rev. Lett. 66, 2712 (1991);
%
A.~Baldit et al.,
Phys. Lett. B 332, 244 (1994);
%
E.A.~Hawker et al.,
Phys. Rev. Lett. 80, 3715 (1998);
%
J.-C.~Peng and G.T.~Garvey,
LA-UR-99-5003, to appear in ``Trends in Particle and Nuclear Physics'',
Plenum Press, New York, hep-ph/9912370.

\bibitem{EARLY}
E.M.~Henley and G.A.~Miller,
Phys. Lett. B 251, 453 (1990);
%
A.I.~Signal, A.W.~Schreiber and A.W.~Thomas,
Mod. Phys. Lett. A 6, 271 (1991);
%
S.~Kumano,
Phys. Rev. D43, 3067 (1991);
%
S.~Kumano and J.T.~Londergan,
Phys. Rev. D 44, 717 (1991);
%
W.-Y.P.~Hwang, J.~Speth and G.E.~Brown,
Z. Phys. A339, 383 (1991).

\bibitem{CLOUD}
For reviews see: 
J.~Speth and A.W.~Thomas,
Adv. Nucl. Phys. 24, 83 (1998);
%
S.~Kumano,
Phys. Rep. 303, 183 (1998).

\bibitem{CHIPT}
S.~Weinberg,
Physica (Amsterdam) 96 A, 327 (1979);
%
J.~Gasser and H.~Leutwyler,
Ann. Phys. 158, 142 (1984);
%
E.~Jenkins, M.~Luke, A.V.~Manohar and M.J.~Savage,
Phys. Lett. B 302, 482 (1993);
%
V.~Bernard, N.~Kaiser and U.-G.~Mei\ss ner,
Int. J. Mod. Phys. E 4, 193 (1995).

\bibitem{SULL}
J.D.~Sullivan,
Phys. Rev. D 5, 1732 (1972).

\bibitem{CBM}
S.~Th\'eberge, G.A.~Miller and A.W.~Thomas,
Phys. Rev. D 22, 2838 (1980);  
%
A.W.~Thomas,
Adv. Nucl. Phys. 13, 1 (1984).

\bibitem{MST}
W.~Melnitchouk, J.~Speth and A.W.~Thomas,
Phys. Rev. D 59, 014033 (1999);
%
F.~Carvalho, F.O.~Durces, F.S.~Navarra and M.~Nielsen,
Phys. Rev. D 60 094015 (1999).

\bibitem{HEAVY}
W.~Melnitchouk and A.W.~Thomas,
Phys. Rev. D 47, 3794 (1993);
%
H.~Holtmann, A.~Szczurek and J.~Speth,
Nucl. Phys. A 596, 631 (1996);
%
M.~Alberg, E.M.~Henley and G.A.~Miller,
Phys. Lett. B 471, 396 (2000).

\bibitem{ST}
A.I.~Signal and A.W.~Thomas,
Phys. Lett. B 191, 206 (1987).

\bibitem{STRANGE}
X.~Ji and J.~Tang,
Phys. Lett. B 362, 182 (1995);
%
S.J.~Brodsky and B.Q.~Ma,
Phys. Lett. B 381, 317 (1996);
%
P.~Geiger and N.~Isgur,   
Phys. Rev. D 55, 299 (1997);
%
W.~Melnitchouk and M.~Malheiro,
Phys. Rev. C 55, 431 (1997);
%
F.~Carvalho, F.O.~Durces, F.S.~Navarra, M.~Nielsen and F.M.~Steffens,
hep-ph/9912378.

\bibitem{CCFR}
A.O.~Bazarko et al.,
Z. Phys. C 65, 189 (1995).

\bibitem{PAULI}
R.D.~Field and R.P.~Feynman,
Phys. Rev. D 15, 2590 (1977);
%
A.W.~Schreiber, A.I.~Signal and A.W.~Thomas,
Phys. Rev. D 44, 2653 (1991);
%
F.M.~Steffens and A.W.~Thomas,
Phys. Rev. C 55, 900 (1997);
%
B.~Dressler, K.~Goeke, M.V.~Polyakov and C.~Weiss,
Eur. Phys. J. C 14, 147 (2000).
\end{document}